\documentclass[preprint,aps,showpacs,floatfix,nofootinbib]{revtex4}
\usepackage{amsfonts}
\usepackage{amsmath}
\usepackage{euscript}
\usepackage{graphicx}
\usepackage{subfigure}
\usepackage{axodraw}

\newcommand{\vev}[1]{\left\langle #1 \right\rangle}

\newcommand{\diag}{\mathop{\rm diag}\nolimits}
\begin{document}
\preprint{VPI--IPPAP--06--03}

\title{Lifetimes of the Heavy Neutral Leptons in the Okamura Model}
\author{Alexey Pronin}\email{apronin@vt.edu}
\author{Tatsu Takeuchi}\email{takeuchi@vt.edu}
\affiliation{Department of Physics, Virginia Tech, Blacksburg, VA 24061}

\date{\today}

\begin{abstract}
We study the lifetimes of TeV-scale heavy neutral leptons (Majorana neutrinos) that appear
in a model suggested by Okamura et al.~\cite{okamura}. 
We develop a convenient way to parametrize the neutrino mass texture of the model, and illustrate our method by calculating the mass spectrum, decay widths, and lifetimes of the heavy particles over the entire parameter space.
From the mass spectrum, we find that for most of the parameter space, only two-body decays are relevant in the calculation of the
lifetime, with typical values falling in the range of $10^{-26}$ to $10^{-24}$ seconds. 
If the particles discussed here are created at colliders, their lifetimes are short enough
for them to decay inside the detector, while long enough to lead to a narrow peak
in the invariant mass spectrum of the decay products. 
However, an analysis by Dicus, Karatas, and Roy \cite{Dicus:1991fk} suggests that
they may be difficult to observe at the LHC.
\end{abstract}


\maketitle
\section{Introduction.}

Several models of neutrino mass have been suggested in the literature in which
the neutrinos acquire masses through a seesaw \cite{seesaw} type mass texture, 
but the Majorana masses of the right-handed neutrinos are at the TeV scale
instead of the GUT scale of $\sim 10^{16}$~GeV \cite{okamura,Chang:1994hz,Carone:1996ny,Arkani-Hamed:2000bq,Ma:2000cc}.
The smallness of the neutrino masses in those models is achieved either
by the reduction of the rank of the mass matrix through a judicious choice of mass texture \cite{okamura,Chang:1994hz}, or by the suppression of the Dirac masses through an extended Higgs sector \cite{Carone:1996ny,Arkani-Hamed:2000bq,Ma:2000cc}.

In such models, the heavy, mostly-right-handed mass eigenstates typically
have masses of a few TeV, placing them within reach 
of the CERN Large Hadron Collider (LHC) or future $e^+e^-$ linear colliders.
If created, the particles will decay into a light neutrino+Higgs through 
the Yukawa interaction responsible for the Dirac masses, 
or into a light neutrino+$Z$ or a charged lepton+$W$ through 
the small admixture of the left-handed neutrino state.
This last decay mode is particularly interesting since the decay products
can be all visible.  
Of course, whether such a decay, and thus the particle, 
can be observed at colliders or not depends 
on whether the lifetime of the particle is short enough for it to decay 
inside the detector, and if that is the case, whether the width is small enough so that a narrow peak is discernible in the invariant mass of its decay products.

In this paper, we calculate the lifetimes of the heavy, mostly-right-handed states
of the model proposed by Okamura et al. in Ref.~\cite{okamura}.
The original motivation of the model was to explain the NuTeV anomaly 
\cite{Zeller:2001hh,Davidson:2001ji}, one possible solution of which requires 
largish mixing ($\theta^2\sim 0.003$) between the light and heavy ($\gg M_Z$) neutrino states 
\cite{nutev}. 
Denoting the left- and right-handed neutrino states by $\nu$ and $\xi$, respectively,
the Okamura texture is given by
\begin{equation}
\left[\;
\overline{\nu^c_1}\;\;
\overline{\nu^c_2}\;\;
\overline{\nu^c_3}\;\;
\overline{\xi}_1\;\;
\overline{\xi}_2\;\;
\overline{\xi}_3\;
\right]
\left[ \begin{array}{cccccc}
        0 &  0 &  0 & \alpha m & \beta m & \gamma m \\
        0 &  0 &  0 & \alpha m & \beta m & \gamma m \\
        0 &  0 &  0 & \alpha m & \beta m & \gamma m \\
        \alpha m & \alpha m & \alpha m & \alpha M & 0 & 0 \\
        \beta  m & \beta  m & \beta  m & 0 & \beta  M & 0 \\
        \gamma m & \gamma m & \gamma m & 0 & 0 & \gamma M \\
        \end{array}
\right]
\left[ \begin{array}{c} \nu_1 \\ \nu_2 \\ \nu_3 \\
                        \xi^c_1 \\ \xi^c_2 \\ \xi^c_3
       \end{array}
\right]\;,
\label{threegentex}
\end{equation}
where the dimensionless parameters $\alpha$, $\beta$, and $\gamma$ are in general complex
and assumed to satisfy the relation
\begin{equation}
\alpha + \beta + \gamma = 0\;.
\label{sum}
\end{equation}
This condition reduces the rank of the above mass matrix to three, leading automatically 
to three massless neutrino states.  Though the actual light, mostly left-handed neutrino states in nature are not completely massless, this model suffices as a first approximation.
We fix the normalization of the three complex parameters $\alpha$, $\beta$, and $\gamma$ to
\begin{equation}
|\alpha|^2+|\beta|^2+|\gamma|^2=3\;.
\label{normalization}
\end{equation}
The dimensionful parameters $m$ and $M$ can be taken to be real and
they set the scale of the Dirac and Majorana masses, 
respectively. The solution to the NuTeV anomaly requires their ratio to be 
\cite{nutev,okamura}
\begin{equation}
\frac{m}{M} \sim 0.03\;.
\label{mMratio}
\end{equation}
If the gauge singlet states $\xi_i$ $(i=1,2,3)$ couple to other particles 
only through the Yukawa interactions which generate the Dirac submatrix of 
Eq.~(\ref{threegentex}), then any permutation of the three complex parameters 
$\alpha$, $\beta$ and $\gamma$ leads to the exact same model since we will have 
the freedom to relabel the three gauge singlet states without affecting any physics.  
In those cases, there exist a $3!=6$ fold redundancy in the parameter space spanned by 
$\alpha$, $\beta$, and $\gamma$.
This will be assumed in the following.

If we set $M=0$ in Eq.~(\ref{threegentex}), we obtain
\begin{equation}
\left[ \begin{array}{cccccc}
        0 & 0 & 0 & \alpha m & \beta m & \gamma m \\
        0 & 0 & 0 & \alpha m & \beta m & \gamma m \\
        0 & 0 & 0 & \alpha m & \beta m & \gamma m \\
        \alpha m & \alpha m & \alpha m & 0 & 0 & 0 \\
        \beta  m & \beta  m & \beta  m & 0 & 0 & 0 \\
        \gamma m & \gamma m & \gamma m & 0 & 0 & 0
        \end{array}
\right]\;
\label{unificationtexture}
\end{equation}
which is manifestly rank 2.  The non-zero eigenvalues of this matrix are
\footnote{A factor of $\sqrt{3}$ is missing from Eq.~(65) of Ref.~\cite{okamura}.}
\begin{equation}
\pm m\sqrt{ 3\left(|\alpha|^2+|\beta|^2+|\gamma|^2 \right)} = \pm 3\,m\;.
\end{equation}
Therefore, this mass texture leads to four massless and two massive Majorana fermions. 
Pairing up the Majorana fermions with the same mass and opposite CP, we can reduce the 
set to one massive and two massless Dirac fermions \cite{bilenky}. 
If we assume that the up-type quarks share the same Dirac mass texture as the
neutrinos, as would be the case in the Pati-Salam model \cite{Pati:uk}, 
we obtain one massive quark which can be identified with the $t$, and 
two massless quarks which can be identified with the $u$ and the $c$.
To produce the $t$ quark mass, we need
\begin{equation}
m\sim 60\,\mathrm{GeV}\;,
\label{mvalue}
\end{equation}
which together with Eq.~(\ref{mMratio}) implies
\begin{equation}
M\sim 2\,\mathrm{TeV}\;.
\label{Mvalue}
\end{equation}
Fixing $m$ and $M$ to these values, the parameter space of the Okamura model
is given by the values of $\alpha$, $\beta$, and $\gamma$ which
satisfy Eqs.~(\ref{sum}) and (\ref{normalization}).

In the following, we introduce a convenient graphical representation of the
parameter space for the Okamura model, and calculate the masses and lifetimes of 
the three heavy mass eigenstates over it.
We find that except for the vicinity of three isolated points at 
the `edge' of the parameter space, 
the three masses are always in the TeV range, and the lifetimes are
typically in the range of $10^{-26}$ to $10^{-24}$ seconds.  In terms of the
widths, these correspond to the range of $0.7\sim 70$~GeV, which are fairly narrow
compared to the masses.

\section{The Parameter Space of the Okamura Model}
\label{symmetry} 

\begin{figure}[ht]
\begin{picture}(500,320)(-30,-250)
\SetScale{1}
\SetWidth{0.5}
\LongArrow(-50,0)(90,0)
\LongArrow(0,-60)(0,60)
\DashLine(0,0)(45,40){3}
\DashLine(70,0)(45,40){3}
\SetWidth{1}
\LongArrow(0,0)(70,0)
\LongArrow(0,0)(-25,40)
\LongArrow(0,0)(-45,-40)
\Text(104,1)[]{$\mathrm{Re} z$}
\Text(0,70)[]{$\mathrm{Im} z$}
\Text(35,-7)[]{$\alpha$}
\Text(-30,46)[]{$\beta$}
\Text(-50,-46)[]{$\gamma$}
\SetOffset(175,0)
\SetWidth{0.5}
\LongArrow(-50,0)(90,0)
\LongArrow(0,-60)(0,60)
\DashLine(0,0)(45,-40){3}
\DashLine(70,0)(45,-40){3}
\SetWidth{1}
\LongArrow(0,0)(70,0)
\LongArrow(0,0)(-25,-40)
\LongArrow(0,0)(-45,40)
\Text(104,1)[]{$\mathrm{Re} z$}
\Text(0,70)[]{$\mathrm{Im} z$}
\Text(35,-7)[]{$\alpha$}
\Text(-30,-46)[]{$\beta$}
\Text(-50,46)[]{$\gamma$}
\SetOffset(350,0)
\SetWidth{0.5}
\LongArrow(-50,0)(90,0)
\LongArrow(0,-60)(0,60)
\DashLine(0,0)(25,40){3}
\DashLine(70,0)(25,40){3}
\SetWidth{1}
\LongArrow(0,0)(70,0)
\LongArrow(0,0)(-45,40)
\LongArrow(0,0)(-25,-40)
\Text(104,1)[]{$\mathrm{Re} z$}
\Text(0,70)[]{$\mathrm{Im} z$}
\Text(35,-7)[]{$\alpha$}
\Text(-50,46)[]{$\beta$}
\Text(-30,-46)[]{$\gamma$}
\SetOffset(0,-140)
\LongArrow(0,0)(70,0)
\LongArrow(70,0)(45,40)
\LongArrow(45,40)(0,0)
\Text(35,-7)[]{$\alpha$}
\Text(65,26)[]{$\beta$}
\Text(15,26)[]{$\gamma$}
\SetOffset(175,-100)
\LongArrow(0,0)(70,0)
\LongArrow(70,0)(45,-40)
\LongArrow(45,-40)(0,0)
\Text(35,+7)[]{$\alpha$}
\Text(65,-26)[]{$\beta$}
\Text(15,-26)[]{$\gamma$}
%
\SetOffset(350,-140)
\LongArrow(0,0)(70,0)
\LongArrow(70,0)(25,40)
\LongArrow(25,40)(0,0)
\Text(35,-7)[]{$\alpha$}
\Text(55,26)[]{$\beta$}
\Text(5,26)[]{$\gamma$}
\SetOffset(0,-240)
\Line(0,0)(70,0)
\Line(70,0)(45,40)
\Line(45,40)(0,0)
\Text(35,-10)[]{$|\alpha|$}
\Text(66,25)[]{$|\beta|$}
\Text(14,25)[]{$|\gamma|$}
\Text(40,15)[]{$\odot$}
\SetOffset(175,-200)
\Line(0,0)(70,0)
\Line(70,0)(45,-40)
\Line(45,-40)(0,0)
\Text(35,+10)[]{$|\alpha|$}
\Text(66,-25)[]{$|\beta|$}
\Text(14,-25)[]{$|\gamma|$}
\Text(40,-15)[]{$\otimes$}
\SetOffset(350,-240)
\Line(0,0)(70,0)
\Line(70,0)(25,40)
\Line(25,40)(0,0)
\Text(35,-10)[]{$|\alpha|$}
\Text(56,25)[]{$|\beta|$}
\Text(4,25)[]{$|\gamma|$}
\Text(30,15)[]{$\odot$}
\SetOffset(0,0)
\Text(45,-70)[]{$\Updownarrow$}
\Text(45,-170)[]{$\Updownarrow$}
\Text(220,-70)[]{$\Updownarrow$}
\Text(220,-170)[]{$\Updownarrow$}
\Text(375,-70)[]{$\Updownarrow$}
\Text(375,-170)[]{$\Updownarrow$}
\Text(300,-220)[]{$\Longleftrightarrow$}
\Text(300,-230)[]{equivalent}
\end{picture}
\caption{The three complex numbers $\alpha$, $\beta$ and $\gamma$ satisfying $\alpha+\beta+\gamma=0$ form a close triangle. 
For each choice of the three lengths $|\alpha|$, $|\beta|$, and $|\gamma|$,
there are two possible orientations of the triangle 
($\odot$ and $\otimes$)
which are related by complex conjugation
(1st and 2nd columns).  
However, the $\otimes$ case is equivalent to the $\odot$ case with the lengths
$|\beta|$ and $|\gamma|$ interchanged (2nd and 3rd columns).
}
\label{closed_triangle}
\end{figure}
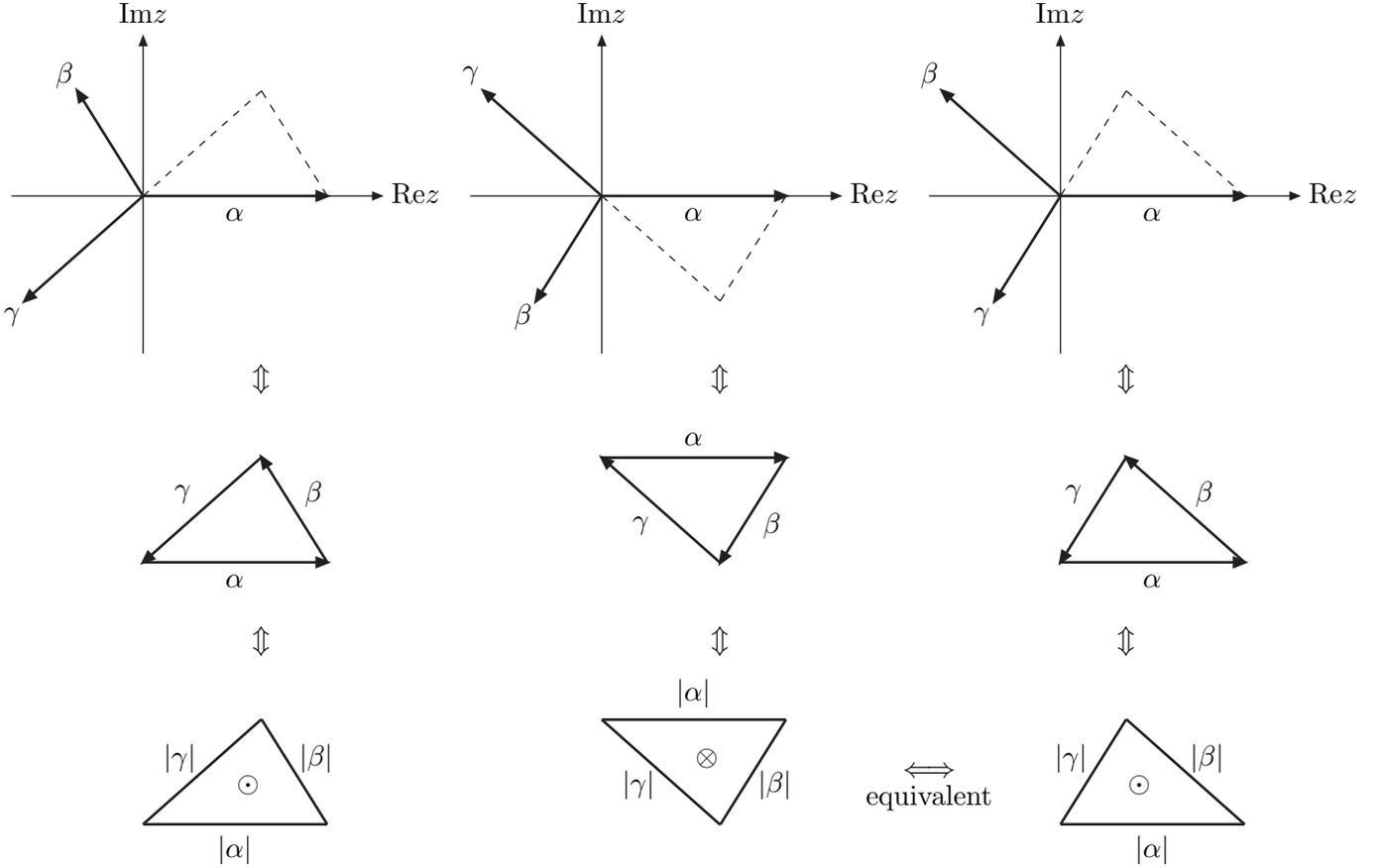

We begin by noting that 
for the three complex parameters $\alpha$, $\beta$, and $\gamma$ to sum to zero,
Eq.~(\ref{sum}), they must form a closed triangle when summed tip-to-tail
as vectors in the complex plane.
Without loss of generality, we can set the phase of $\alpha$ to zero.
This can always be achieved by changing the overall phase of 
$\alpha$, $\beta$ and $\gamma$, and does not affect any physical result.
Therefore, the triangle formed by $\alpha$, $\beta$, and $\gamma$ 
can be assumed to have its base along the positive real axis. 
We define the ``orientation'' of this triangle as the direction of the 
vectorial cross product $\alpha\times\beta$.
If the orientation of the triangle is $\odot$ (out of the complex plane),
then $\beta$ is in the upper complex plane while $\gamma$ is in the lower complex plane.
If the orientation of the triangle is $\otimes$ (into the complex plane),
then $\beta$ is in the lower complex plane while $\gamma$ is in the upper complex plane
(see Fig.~\ref{closed_triangle}).
Then, it is easy to see that
specifying the lengths of the three sides $|\alpha|$, $|\beta|$, and $|\gamma|$, 
and the orientation of the triangle is equivalent to specifying the three complex
numbers $\alpha$, $\beta$, and $\gamma$.

Furthermore, we need not consider both orientations since the two cases 
can be transformed into each other by a simple interchange
of the lengths of $\beta$ and $\gamma$, and a relabeling of the singlet neutrino fields.
As discussed previously, this does not affect any physical result either.
Therefore, we will always take the triangle to be in the $\odot$ orientation.
This choice also reduces the redundancy of the parameter space
from $3!=6$ to 3 since we have used up the freedom to interchange $\beta$ and $\gamma$
to fix the orientation.  

\begin{figure}[t]
\begin{picture}(300,250)(-150,-100)
\SetWidth{1}
\BCirc(0,0){50}
\SetWidth{0.5}
\LongArrow(-150,0)(150,0)
\LongArrow(0,-100)(0,150)
\Line(-86.6,-50)(86.6,-50)
\Line(-86.6,-50)(0,100)
\Line(86.6,-50)(0,100)
\Vertex(10,20){2}
\DashLine(10,20)(10,-50){3}
\DashLine(10,20)(37.14,35.67){3}
\DashLine(10,20)(-32.14,44.33){3}
\Text(22,-20)[]{$|\alpha|^2$}
\Text(32,20)[]{$|\beta|^2$}
\Text(-20,22)[]{$|\gamma|^2$}
\Text(-5,105)[]{2}
\Text(-5,57)[]{1}
\Text(-5,-7)[]{0}
\Text(-7,-57)[]{-1}
\Text(-56,-7)[]{-1}
\Text(45,-7)[]{1}
\Vertex(0,0){1.5}
\Vertex(-50,0){1.5}
\Vertex(50,0){1.5}
\Vertex(0,50){1.5}
\Vertex(0,100){1.5}
\Vertex(0,-50){1.5}
\Vertex(-86.6,-50){1.5}
\Vertex(86.6,-50){1.5}
\end{picture}
\caption{The distances from any point inside an equilateral triangle of height three 
to its three sides add up to three. We can use these distances to
specify $|\alpha|^2$, $|\beta|^2$, and $|\gamma|^2$.  
The triangle inequality is satisfied for points inside the unit circle which
inscribes the triangle.}
\label{triangle-circle}
\end{figure}
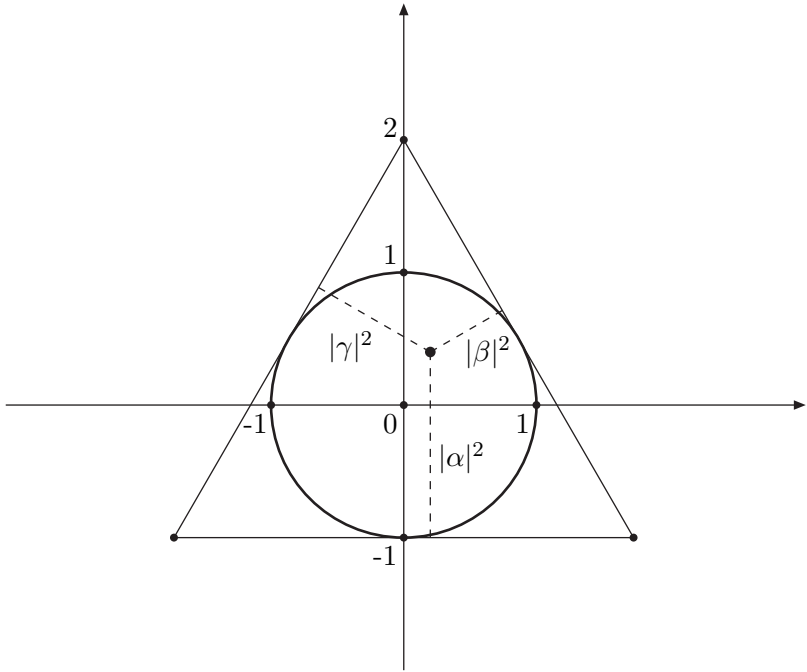

This consideration shows that specifying the three lengths $|\alpha|$, $|\beta|$,
and $|\gamma|$ suffices to uniquely determine the Okamura texture, with
cyclic permutations of the three lengths leading to the same model.
(This residual redundancy comes from our freedom to 
choose which of the three lengths to call $|\alpha|$.)
The question then, is, how can we specify those three lengths so they satisfy the normalization condition Eq.~(\ref{normalization}), and also 
the triangle inequalities:
\begin{equation} \label{trianglecondition}
	        |\beta|+|\gamma|\ge|\alpha|\;,\quad    
	        |\alpha|+|\beta|\ge|\gamma|\;, \quad  
	        |\alpha|+|\gamma|\ge |\beta|\;, 		  
\end{equation}
so they form a closed triangle?
To this end, we utilize the fact that the sum of distances from any point inside
a triangle to its three sides is constant: any point inside an equilateral triangle of
height three will have distances to the three sides which add up to three.  
If we identify these distances with $|\alpha|^2$, $|\beta|^2$, and
$|\gamma|^2$, we can use the position of the point to specify the three lengths.
Requiring the square-roots of these distances to satisfy the
triangle inequality constrains the point to be
inside a unit circle which inscribes the triangle.
Therefore, for every point inside the unit circle, we can associate a corresponding
parameter set for the Okamura texture (see Fig.~\ref{triangle-circle}).

\begin{figure}[ht]
\includegraphics[width=5cm]{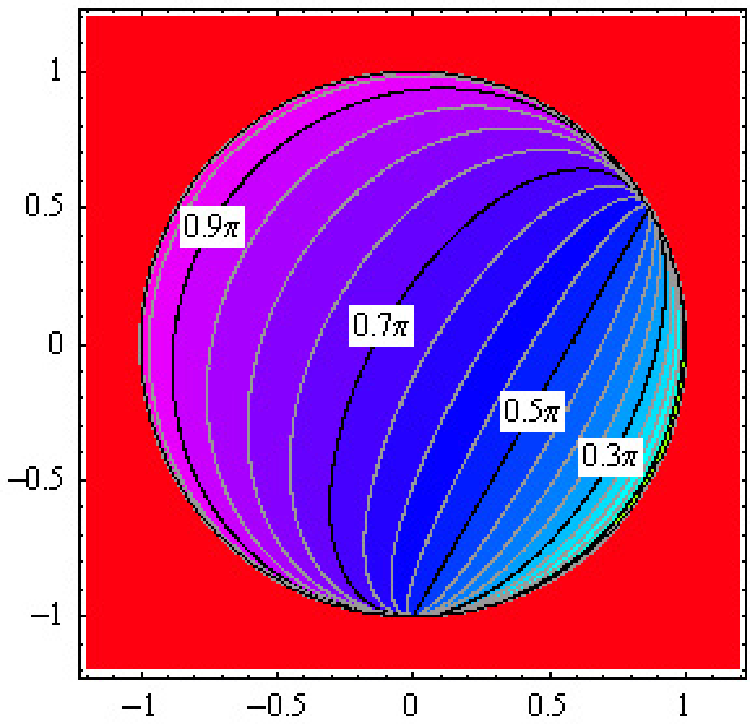}
\includegraphics[width=5cm]{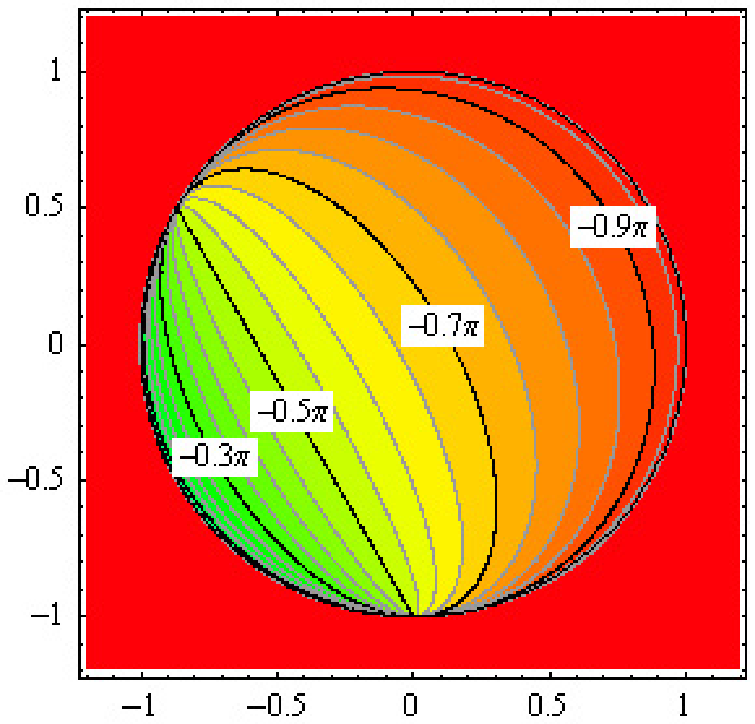}
\includegraphics[height=5.4cm]{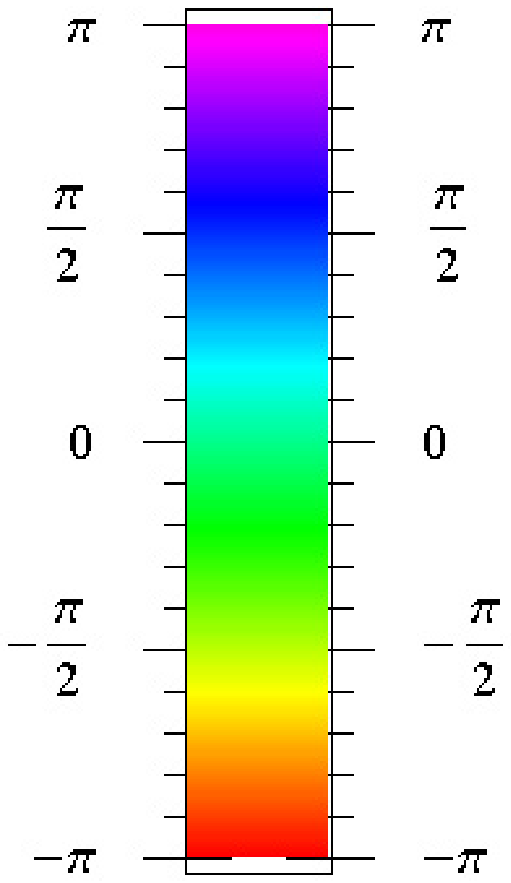}
\caption{Contour lines and density plots for $\arg\beta$ (left) and $\arg\gamma$ (middle). The distances between two consecutive equipotential lines are $\Delta\arg\beta=\Delta\arg\gamma=0.05\pi$. The color scheme is shown on the right.}
\label{phases}
\end{figure}

If we specify the position of a point inside the unit circle with its 
polar coordinate $(r,\theta)$, where $0 \le r \le 1$, $\theta\in [-\pi,\pi)$, 
the corresponding values of $|\alpha|$, $|\beta|$, and $|\gamma|$ are:
\begin{equation}	
|\alpha| = \sqrt{1+r\sin\theta}\;,\quad 
|\beta|  = \sqrt{1+r\sin \left(\theta-\frac{2\pi}{3}\right)}\;,\quad
|\gamma| = \sqrt{1+r\sin \left(\theta+\frac{2\pi}{3}\right)}\;.
\label{abg-rt}
\end{equation}
The phases of the three numbers are:
\begin{eqnarray} \label{abg-phases}
\arg\alpha & = & \phantom{-}0\;,\cr
\arg\beta  & = & \phantom{-}\pi
-\cos^{-1}\frac{1/2+r\sin(\theta-\pi/3)}
               {\sqrt{[\,1+r\sin\theta\,][\,1+r\sin (\theta-2\pi/3)\,]}}\;,\cr
\arg\gamma & = & -\pi
+\cos^{-1}\frac{1/2+r\sin(\theta+\pi/3)}
               {\sqrt{[\,1+r\sin\theta\,][\,1+r\sin (\theta+2\pi/3)\,]}}\;.	
\end{eqnarray}
In Fig.~\ref{phases}, we plot the dependence of $\arg\beta$ and $\arg\gamma$
on the position of the point inside the unit circle.

A cyclic permutation of $\alpha$, $\beta$, and $\gamma$ which leaves the physics invariant up to an overall phase corresponds to the transformation $\theta\rightarrow\theta+2\pi/3$ ($120^\circ$ rotations). 
This means that we expect the same symmetry to be present in the mass spectrum and 
the values of heavy neutrino decay widths and lifetimes. 
This can be used as a useful check of our calculations.

\section{The Lagrangian}

To calculate the lifetimes of the heavy neutral states, we must first specify their
interactions.
We denote the left-handed charged lepton fields with $\ell$,
and the left- and right-handed neutrino fields with $\nu$ and $\xi$,
respectively:
\begin{equation}
\ell  = \left[ \begin{array}{c} \ell_1 \\ \ell_2 \\ \ell_3 \end{array} \right] \;,\qquad
\nu   = \left[ \begin{array}{c} \nu_1  \\ \nu_2  \\ \nu_3  \end{array} \right] \;,\qquad
\xi = \left[ \begin{array}{c} \xi_1 \\ \xi_2 \\ \xi_3 \end{array} \right] \;.
\end{equation}
The right-handed neutrino fields, $\xi_i$ ($i=1,2,3$), are gauge singlets.
The components of the Higgs doublet are denoted
\begin{eqnarray}
H = \left[ \begin{array}{c} \phi^+ \\ \phi_0 \end{array} \right] \;.
\end{eqnarray}
Then, the Lagrangian which governs the interaction of the neutrinos is
\begin{equation}
\mathcal{L} = \mathcal{L}_{W,Z} + \mathcal{L}_{H} + \mathcal{L}_{M}\;,
\end{equation}
where
\begin{eqnarray}
\mathcal{L}_{W,Z} & = & 
  \frac{g}{\sqrt{2}}\left( \overline{\ell}\gamma^\mu \nu \right) W^{-}_\mu
+ \frac{g}{\sqrt{2}}\left( \overline{\nu}\gamma^\mu \ell \right) W^{+}_\mu 
+ \frac{g}{2\cos\theta_W}\left(\overline{\nu}\gamma^\mu \nu\right) Z_\mu \;,\cr
\mathcal{L}_{H} & = &  
- \overline{\xi}\, \lambda \left( \phi^0 \nu - \phi^{+} \ell \right)
+ h.c. \;,\cr
\mathcal{L}_{M} & = &
-\frac{1}{2}\,\overline{\xi}\, \mathcal{M}\, \xi^c + h.c. \;.
\end{eqnarray}
We neglect the Yukawa interactions which give rise to the charged lepton masses:
the charged leptons are treated as massless as well as the light neutrino states.
In the Okamura model, the Yukawa matrix $\lambda$ and the Majorana mass matrix
$\mathcal{M}$ are
given by
\begin{equation}
\lambda = \frac{\sqrt{2}{m}}{v}
\left[ \begin{array}{ccc} 
\alpha & \alpha & \alpha \\
\beta  & \beta  & \beta  \\
\gamma & \gamma & \gamma 
\end{array} \right]\;,\qquad
\mathcal{M} = M
\left[ \begin{array}{ccc}
\alpha & 0 & 0 \\
0 & \beta & 0 \\
0 & 0 & \gamma
\end{array} \right]\;.
\end{equation}
After the neutral Higgs develops a VEV,
\begin{equation}
\vev{\phi^0} = \vev{\phi^{0*}} = \frac{v}{\sqrt{2}}\;,
\end{equation}
the Yukawa matrix $\lambda$ leads to the Dirac mass matrix of the neutrinos:
\begin{equation}
\mathcal{D} = \frac{v}{\sqrt{2}}\lambda
= m \left[ \begin{array}{ccc}
\alpha & \alpha & \alpha \\
\beta  & \beta  & \beta  \\
\gamma & \gamma & \gamma
\end{array} \right] \;.
\end{equation}
The Goldstone bosons are absorbed into the $W$ and the $Z$, $\phi^0\rightarrow 1/\sqrt{2}(h+v)$ as usual, and the resulting
Lagrangian is:
\begin{eqnarray}
\mathcal{L} & = & 
  \frac{g}{\sqrt{2}}\left( \overline{\ell}\gamma^\mu \nu  \right) W^{-}_\mu
+ \frac{g}{\sqrt{2}}\left( \overline{\nu} \gamma^\mu \ell \right) W^{+}_\mu 
+ \frac{g}{2\cos\theta_W}\left( \overline{\nu}\gamma^\mu \nu \right) Z_\mu \cr
& &
- \overline{\xi}\,\mathcal{D}\,\nu
- \frac{1}{\sqrt{2}}\left(\overline{\xi}\,\lambda\,\nu \right) h
- \frac{1}{2}\,\overline{\xi}\, \mathcal{M}\, \xi^c + h.c.
\end{eqnarray}
The neutrino mass terms can be written as
\begin{eqnarray}
\lefteqn{\overline{\xi}\,\mathcal{D}\,\nu 
+ \frac{1}{2}\,\overline{\xi}\, \mathcal{M}\, \xi^c  + h.c.} \cr
& = & \frac{1}{2}\left( \overline{\xi}\,\mathcal{D}\,\nu 
+ \overline{\nu^c}\,\mathcal{D}^{T}\,\xi^c
+ \overline{\xi}\,\mathcal{M}\,\xi^c \right) + h.c. \cr
& = & \frac{1}{2}\left[ \begin{array}{cc} \overline{\nu^c} & \overline{\xi} \end{array} \right]
\left[ \begin{array}{cc} 0 & \mathcal{D}^{T} \\ \mathcal{D} & \mathcal{M} \end{array} \right]
\left[ \begin{array}{c} \nu \\ \xi^c \end{array} \right] + h.c.
\end{eqnarray}
This mass matrix is diagonalized with a unitary transformation
involving the $\nu$ and $\xi^c$ fields:
\begin{equation}
\left[ \begin{array}{c} \nu \\ \xi^c \end{array} \right]
= U\left[ \begin{array}{c} \eta \\ \chi \end{array} \right] \;,
\label{diagonalization}
\end{equation}
so that
\begin{equation}
\left[ \begin{array}{cc} \overline{\nu^c} & \overline{\xi} \end{array} \right]
\left[ \begin{array}{cc} 0 & \mathcal{D}^{T} \\ \mathcal{D} & \mathcal{M} \end{array} \right]
\left[ \begin{array}{c} \nu \\ \xi^c \end{array} \right]
= 
\left[ \begin{array}{cc} \overline{\eta^c} & \overline{\chi^c} \end{array} \right]
U^T
\left[ \begin{array}{cc} 0 & \mathcal{D}^{T} \\ \mathcal{D} & \mathcal{M} \end{array} \right]
U
\left[ \begin{array}{c} \eta \\ \chi \end{array} \right]
= 
\left[ \begin{array}{cc} \overline{\eta^c} & \overline{\chi^c} \end{array} \right] 
M_\mathrm{diag}
\left[ \begin{array}{c} \eta \\ \chi \end{array} \right] \;,
\end{equation}
with $M_\mathrm{diag} = \diag(0,0,0,M_1,M_2,M_3)$.
The $\eta$ and $\chi$ fields are the left-handed mass eigenfields
with $\eta$ being the light (massless) states, and $\chi$ being the
heavy states.
Decomposing the $6\times 6$ matrix $U$ into four $3\times 3$ matrices as
\begin{equation}
U = 
\left[ \begin{array}{cc} U_{\nu\eta} & U_{\nu\chi} \\ 
                         U_{\xi\eta} & U_{\xi\chi}
       \end{array} 
\right]\;,
\end{equation}
we can write
\begin{eqnarray}
\nu & = & U_{\nu\eta}\,\eta + U_{\nu\chi}\,\chi \;, \cr
\xi & = & U_{\xi\eta}^* \eta^c + U_{\xi\chi}^* \chi^c \;,
\end{eqnarray}
(Because the $\eta$ fields are exactly massless and degenerate in our model,
the matrices $U_{\nu\eta}$ and $U_{\xi\eta}$ are not uniquely determined. However,
this does not affect our final results.  Note also, that though $U$ is unitary,
its four $3\times 3$ submatrices are non-unitary in general.)
The relevant interaction terms in the Lagrangian involving the
$\chi$ fields are then:
\begin{eqnarray}
\frac{g}{2\cos\theta_{W}}
\left( \overline{\nu}\gamma^\mu \nu \right) Z_\mu 
& \rightarrow &
\frac{g}{2\cos\theta_{W}}
\Bigl[\, \overline{\eta}\left( U_{\nu\eta}^\dagger U_{\nu\chi}\right) \gamma^\mu \,\chi
      + \overline{\chi}\left( U_{\nu\chi}^\dagger U_{\nu\eta}\right) \gamma^\mu \,\eta
\,\Bigr] Z_\mu \;, 
\cr
\frac{g}{\sqrt{2}}
\left( \overline{\ell}\gamma^\mu \nu \right) W_\mu^- 
& \rightarrow &
\frac{g}{\sqrt{2}}
\left( \overline{\ell}\, U_{\nu\chi}\gamma^\mu \,\chi
\right) W_\mu^- \;, 
\cr 
\frac{1}{\sqrt{2}}
\left(\overline{\xi}\,\lambda\,\nu\right) h 
& \rightarrow &
\frac{1}{\sqrt{2}}
\Bigl[\, \overline{\eta^c}\left( U_{\xi\eta}^{T}\lambda U_{\nu\chi} \right) \chi
       + \overline{\chi^c}\left( U_{\xi\chi}^{T}\lambda U_{\nu\eta} \right) \eta
\,\Bigr] h \;,
\end{eqnarray}
plus the Hermitian conjugates of the later two lines.
Introducing the Majorana fields
\begin{eqnarray}
n = \eta + \eta^c\;,\qquad N = \chi + \chi^c\;,
\end{eqnarray}
(note that these fields do not have definite lepton number)
we can write
\begin{eqnarray}
\eta   = P_L n\;,\qquad
\eta^c = P_R n\;,\qquad
\chi   = P_L N\;,\qquad
\chi^c = P_R N\;,
\end{eqnarray}
and the relevant interaction Lagrangian in terms of these fields becomes
\begin{eqnarray} 
\label{lagrmass}
\mathcal{L} & = &
\frac{g}{2\cos\theta_W}
\Bigl[\,\overline{n}\left(A\gamma^{\mu}P_L-A^*\gamma^{\mu}P_R\right)N\,\Bigr] Z_{\mu} \cr 
& & 
+\, \frac{g}{\sqrt{2}}\left(\overline{\ell}\,B\gamma^{\mu}P_L\,N\right) W^{-}_{\mu} 
- \frac{g}{\sqrt{2}}\left(\overline{\ell^c}\,B^*\gamma^{\mu}P_R\,N\right) W^{+}_{\mu} \cr
& & -\, \overline{n}\left(C h P_L + C^* \tilde h P_R  \right) N\;,
\end{eqnarray}
where 
\begin{equation} \label{abc}
A\equiv U^{\dagger}_{\nu\eta}U_{\nu\chi}\;\text{,}\qquad 
B\equiv U_{\nu\chi}\;\text{,}\qquad
C\equiv \frac{1}{\sqrt{2}}
\left(U^T_{\nu\eta}\lambda^T U_{\xi\chi} + U_{\xi\eta}^T \lambda U_{\nu\chi}\right).
\end{equation}
We have used the generic relations \cite{frules} 
\begin{equation}
\overline{\psi}_1\,O P_{R,L}\,\psi_2
= \overline{\psi^c_2}\, O^T P_{R,L}\,\psi^c_1\;,\quad
\overline{\psi}_1\,O \gamma^{\mu}P_{R,L}\,\psi_2
= -\overline{\psi^c_2}\, O^T \gamma^{\mu}P_{L,R}\,\psi^c_1\;,
\end{equation}
($O$ is a matrix which carries flavor indices only), and 
the fact that $n^c=n$ and $N^c=N$ by construction, 
to rearrange the terms in Eq.~(\ref{lagrmass}) in such a way that all the 
$N$-fields stand at the rightmost position of each term to facilitate the extraction of the
$N$-decay matrix elements.

\section{Lifetimes}

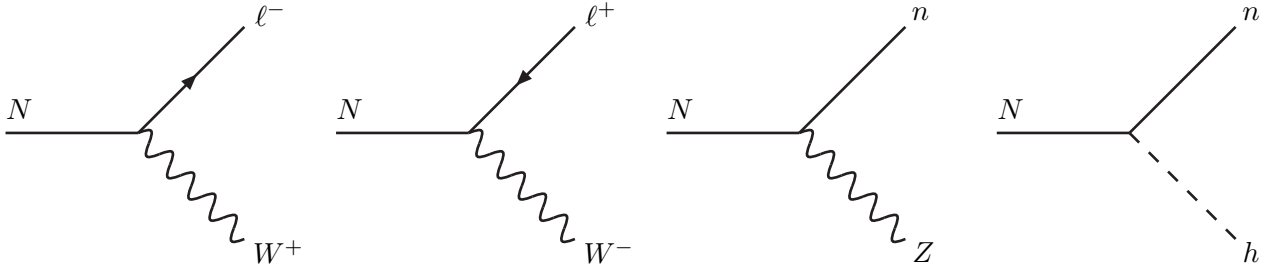
\begin{figure}[ht]
\begin{picture}(500,100)(-50,-50)
\SetWidth{1}
\Line(-50,0)(0,0)
\ArrowLine(0,0)(40,40)
\Photon(0,0)(40,-40){4}{5}
\Text(-45,9)[]{$N$}
\Text(50,45)[]{$\ell^{-}$}
\Text(53,-45)[]{$W^+$}
\SetOffset(125,0)
\Line(-50,0)(0,0)
\ArrowLine(40,40)(0,0)
\Photon(0,0)(40,-40){4}{5}
\Text(-45,9)[]{$N$}
\Text(50,45)[]{$\ell^{+}$}
\Text(53,-45)[]{$W^-$}
\SetOffset(250,0)
\Line(-50,0)(0,0)
\Line(0,0)(40,40)
\Photon(0,0)(40,-40){4}{5}
\Text(-45,9)[]{$N$}
\Text(46,45)[]{$n$}
\Text(47,-45)[]{$Z$}
\SetOffset(375,0)
\Line(-50,0)(0,0)
\Line(0,0)(40,40)
\DashLine(0,0)(40,-40){5}
\Text(-45,9)[]{$N$}
\Text(46,45)[]{$n$}
\Text(46,-45)[]{$h$}
\end{picture}
\caption{The 2-body decay processes of the heavy neutrino $N$.}
\label{processes}
\end{figure}

From Eq.~(\ref{lagrmass}), we can immediately derive the amplitudes 
for the two-body decay processed of the heavy neutrinos, 
$N_i$ ($i=1,2,3$), shown in Fig.~\ref{processes}. 
If the $N_i$ were lighter than the $W$, $Z$, or $h$, then we will need to consider
three-body decay processes mediated by these particles, 
but it turns out that except for small neighborhoods around isolated points 
in the parameter space of the model, they are always heavier.
It therefore suffices to consider only the two-body decay modes.

Now, straightforward calculations allow us to write down the partial decay widths 
for each channel of decay (the indices $i$ and $j$ below run from $1$ to $3$):
\begin{eqnarray} \label{widths}
\Gamma(N_i \rightarrow n_j Z)
& = &\frac{\sqrt{2}G_F|A^{ji}|^2}{16\pi}M_i^3\left(1-\frac{m_Z^2}{M_i^2}\right)^2\left(1+2\frac{m_Z^2}{M_i^2}\right)\;, \cr
\Gamma(N_i \rightarrow \ell_j^{+} W^{-}) 
\;=\; \Gamma(N_i\rightarrow \ell_j^{-} W^{+})
& = & \frac{\sqrt{2}G_F|B^{ji}|^2}{16\pi}M_i^3\left(1-\frac{m_W^2}{M_i^2}\right)^2\left(1+2\frac{m_W^2}{M_i^2}\right)\;, \cr
\Gamma(N_i \rightarrow n_j h)
& = & \frac{|C^{ji}|^2}{16\pi}M_i\left(1-\frac{m_h^2}{M_i^2}\right)^2\;.
\end{eqnarray}
The first two lines can be compared with the results of Djouadi in Ref.~\cite{lifetime}. 
At first sight, these expressions may seem to imply
that the $N\rightarrow n h$ channel is suppressed with respect to the other two
since its partial width grows linearly with the mass $M_i$, while for the 
$N\rightarrow n Z$ and $N\rightarrow \ell W$ channels the widths grow as $M_i^3$. 
However, the interactions of the heavy Majorana neutrino $N$ with the gauge bosons are suppressed because only a small fraction of $N$ is the left-handed neutrino $\nu$. 
Since most of $N$ is the right-handed neutrino $\xi$, no such suppression exists for 
its interaction with the Higgs $h$. 
Numerically, it turns out that all three channels of decay must
be taken into account.


\section{Results}
\label{sec:Results}

\begin{figure} 
	\centering

	\includegraphics[width=5cm]{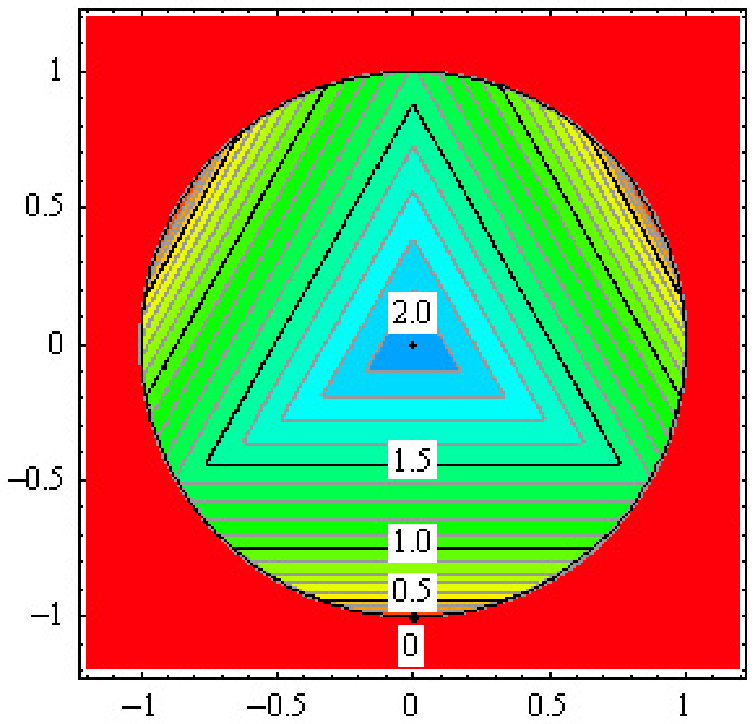}
	\includegraphics[width=5cm]{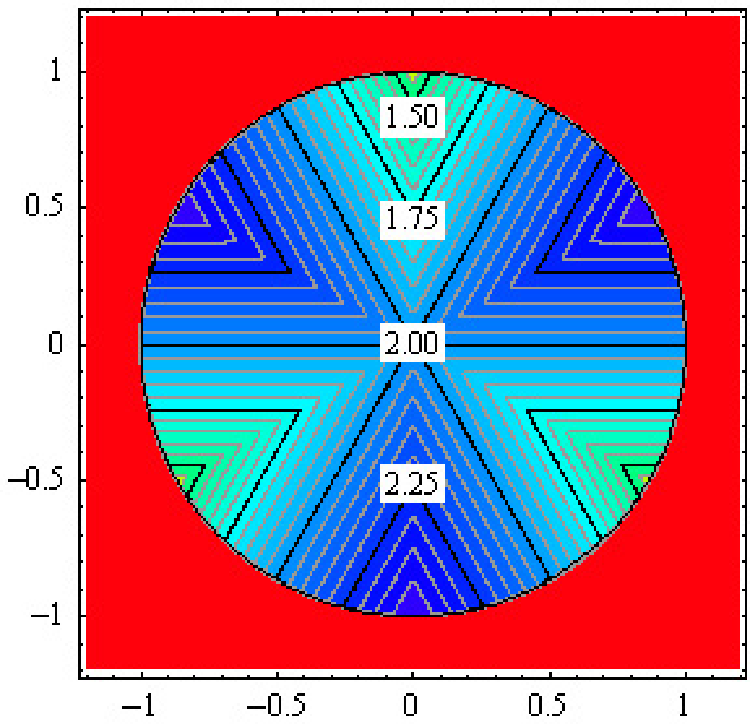}
	\includegraphics[width=5cm]{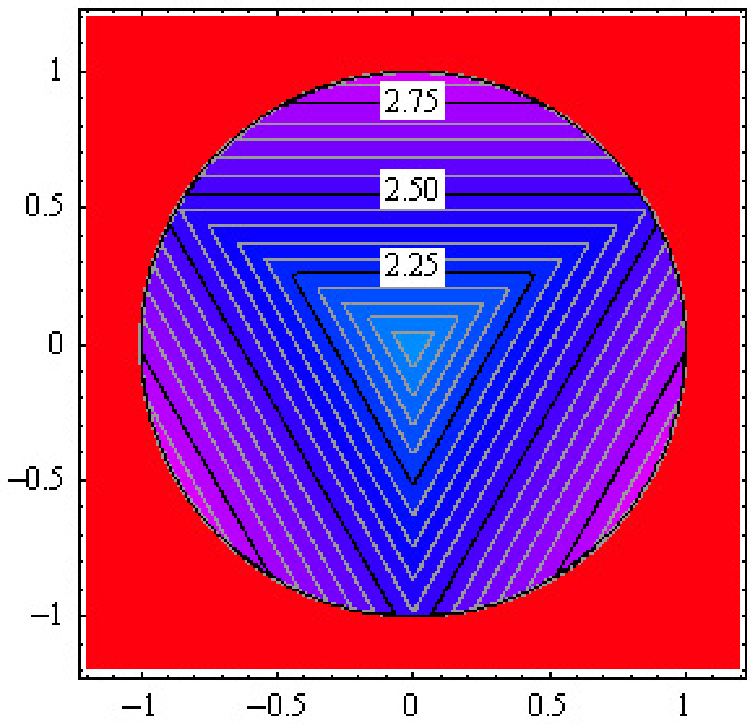}	
	
	\begin{picture}(250,2)(0,0)
	\put(-21,4){(a)}
	\put(125,4){(b)}
	\put(271,4){(c)}
	\end{picture}	
	
	\includegraphics[width=9cm,height=3cm]{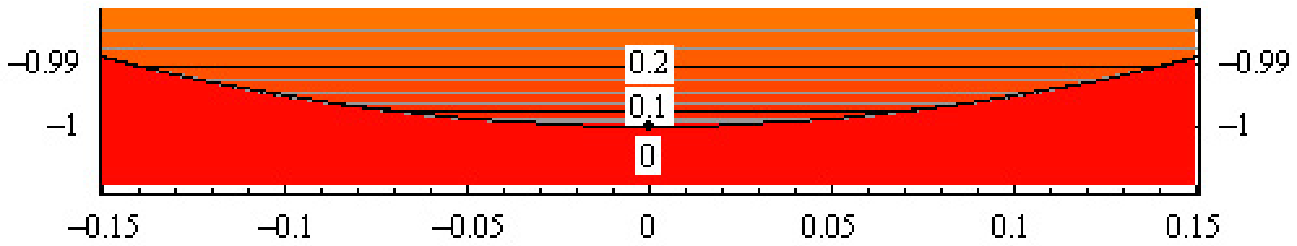}	
	\includegraphics[width=7cm]{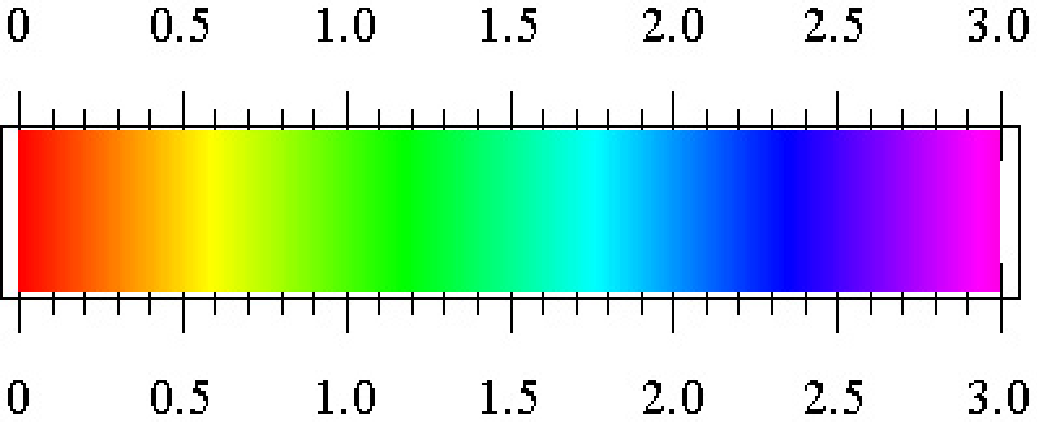}
	
	\begin{picture}(250,2)(0,0)
	\put(15,0){(d)}
	\put(247,0){(e)}
	\end{picture}
	
	\caption{(a), (b), (c) density and contour plots for masses $M_1$, $M_2$, and $M_3$ of the lightest $N_1$, medium heavy $N_2$ and heaviest $N_3$ heavy neutrino respectively (TeV). The distances between two consecutive equipotential lines are $\Delta M_1=0.1$~TeV, $\Delta M_2=\Delta M_3=0.05$~TeV; (d)  the vicinity of the point where $M_1$ approaches zero ($r=1$, $\theta=-\pi/2$); $\Delta\ M_1=0.025$~TeV; (e) mass color coding.}
\label{Mass}	
\end{figure}

Now we have everything at hand to calculate the lifetimes of the heavy neutrinos in 
the Okamura model.
The parameter space of the model is represented by the interior of 
a unit circle as discussed in section~\ref{symmetry}.
For each point inside the unit circle, we can calculate the Okamura texture 
using Eqs.~(\ref{threegentex}), (\ref{mvalue}), (\ref{Mvalue}), (\ref{abg-rt}), 
and (\ref{abg-phases}), diagonalize it to obtain the masses and mixings \cite{algorithm}, 
and calculate the decay widths and
lifetimes of the heavy neutrinos using Eq.~(\ref{widths}).
As the Higgs mass, we take $m_H=200$~GeV. 
(The choice of the Higgs mass has little effect on our result 
as long as $m_H \ll M_i$.)

\begin{figure} 
	\centering
	
	\includegraphics[width=5cm]{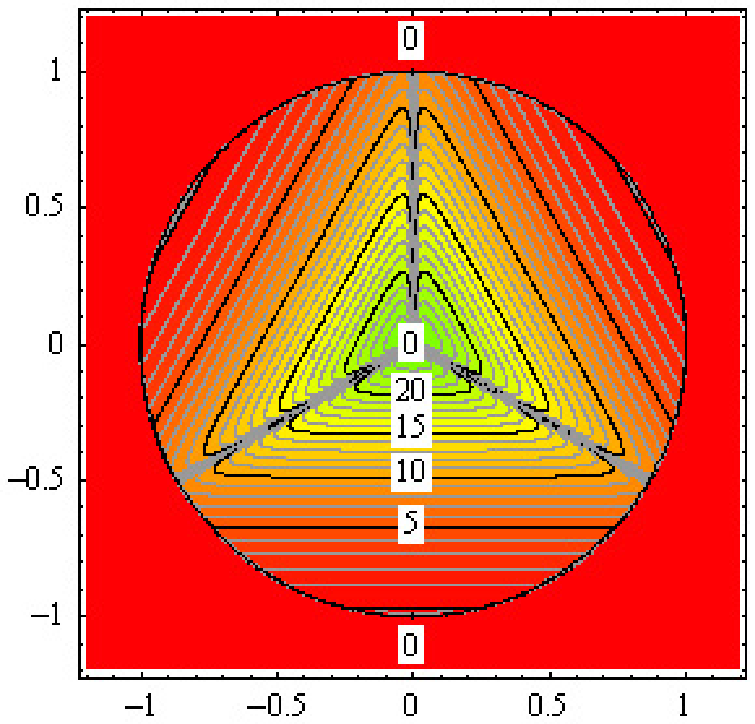}			
	\includegraphics[width=5cm]{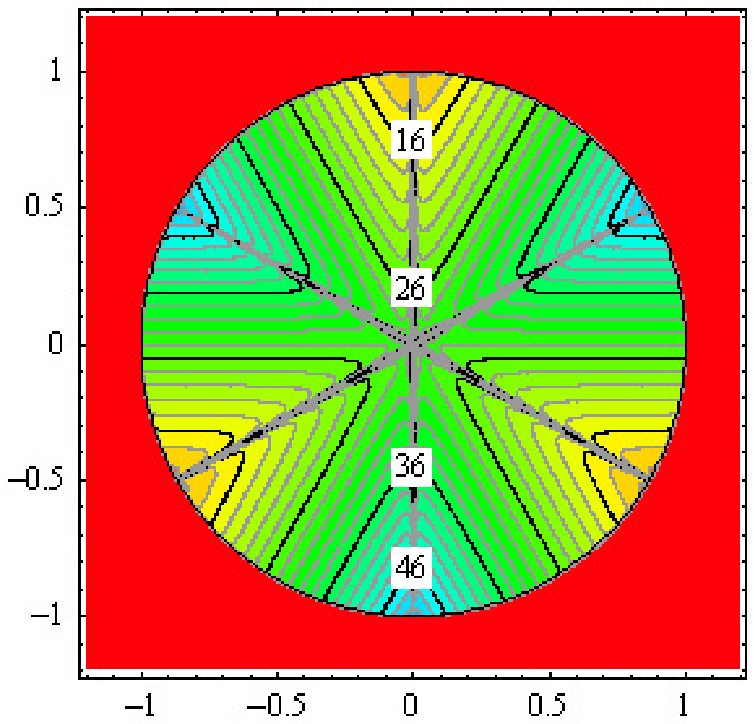}
	\includegraphics[width=5cm]{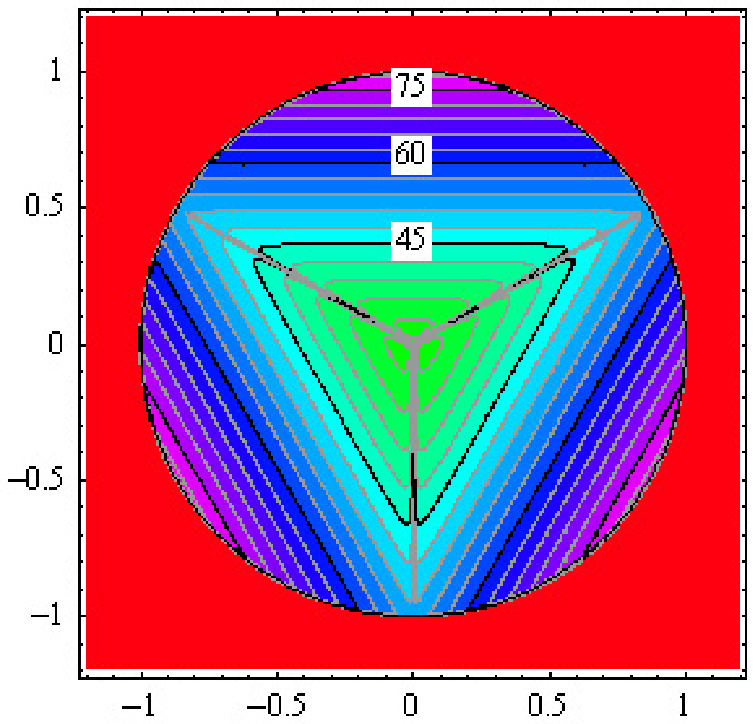}
	
	\begin{picture}(250,2)(0,0)
	\put(-21,4){(a)}
	\put(125,4){(b)}
	\put(271,4){(c)}
	\end{picture}	
	
	\includegraphics[width=5cm]{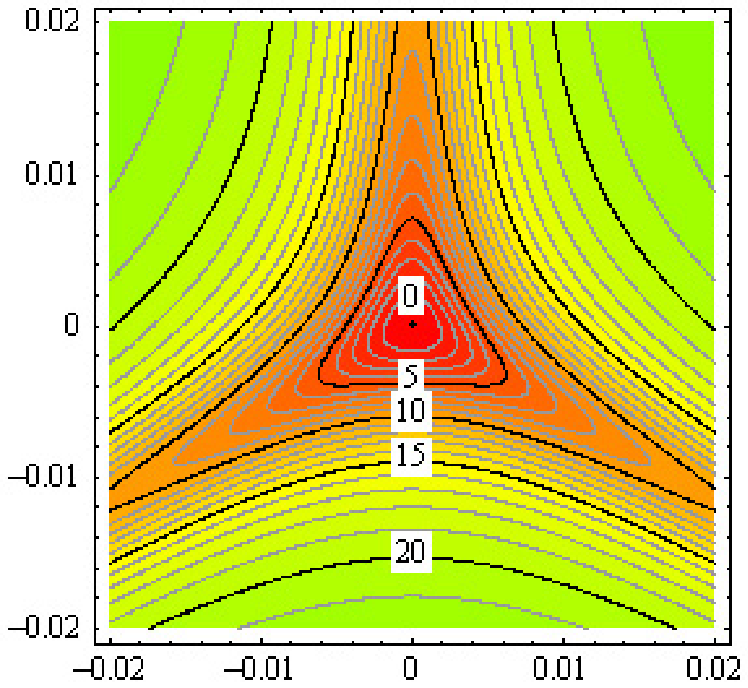}
	\begin{picture}(87,2)(0,0)
	\end{picture}	
	\includegraphics[width=7cm]{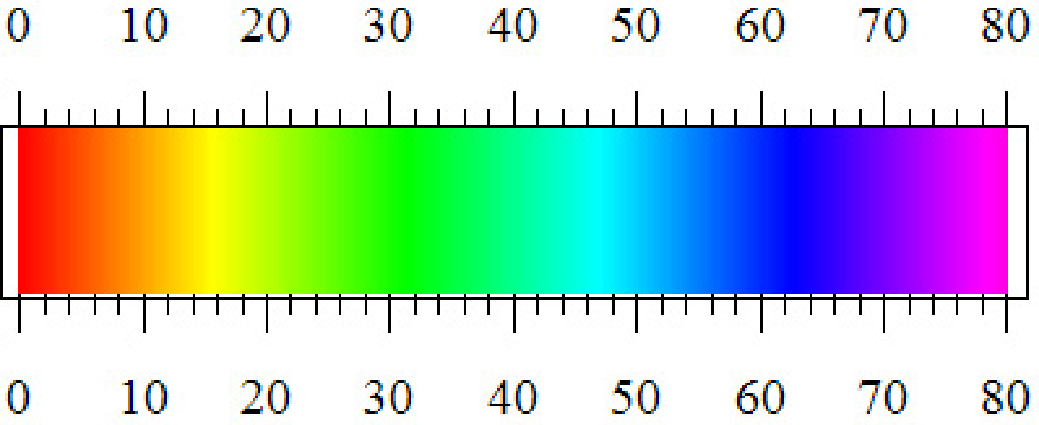}

	\begin{picture}(250,2)(0,0)
  \put(-21,4){(d)}
	\put(234,4){(e)}
	\end{picture}		
			
	\caption{(a), (b), (c) density and contour plots for widths $\Gamma_1$, $\Gamma_2$, and $\Gamma_3$ of the lightest $N_1$, medium heavy $N_2$ and heaviest $N_3$ heavy neutrino respectively (GeV). The distances between two consecutive equipotential lines are $\Delta\Gamma_1=1$~GeV, $\Delta\Gamma_2=2$~GeV, $\Delta\Gamma_3=3$~GeV; (d)
	the detailed picture of the central part of  $\Gamma_1$;
	$\Delta\Gamma_1=
	1$~GeV; 
		(e) 
	width color coding.}  
	
\label{Width}
\end{figure}

\begin{figure} 
	\centering
	\includegraphics[width=5cm]{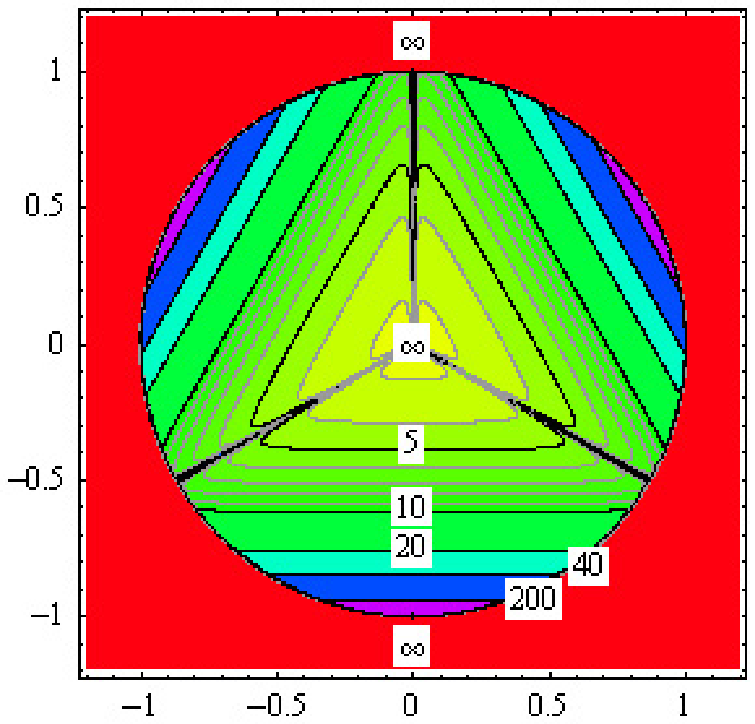}
	\includegraphics[width=5cm]{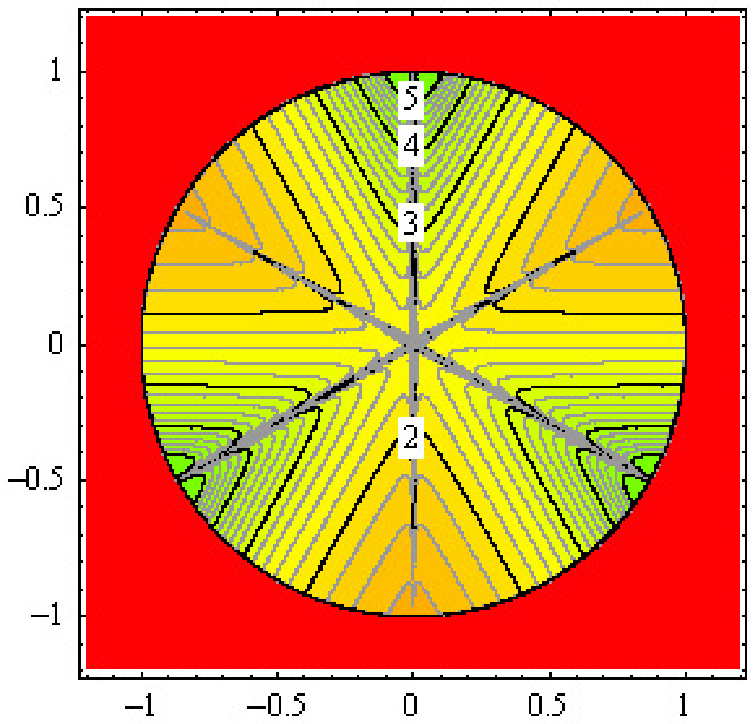}
	\includegraphics[width=5cm]{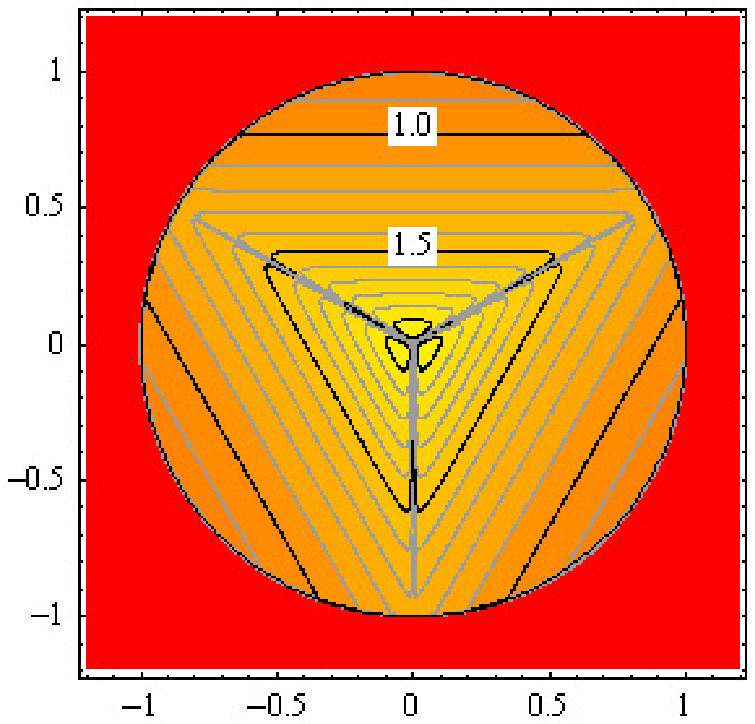}
	
	\begin{picture}(250,2)(0,0)
	\put(-21,4){(a)}
	\put(125,4){(b)}
	\put(271,4){(c)}
	\end{picture}	
	
	\includegraphics[width=5cm]{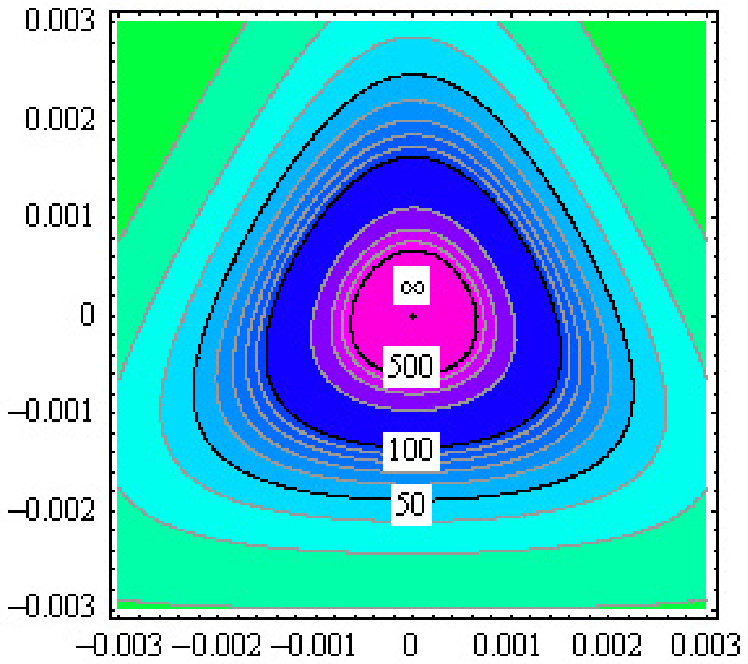}
	\begin{picture}(87,2)(0,0)
	\end{picture}	
	\includegraphics[width=7cm]{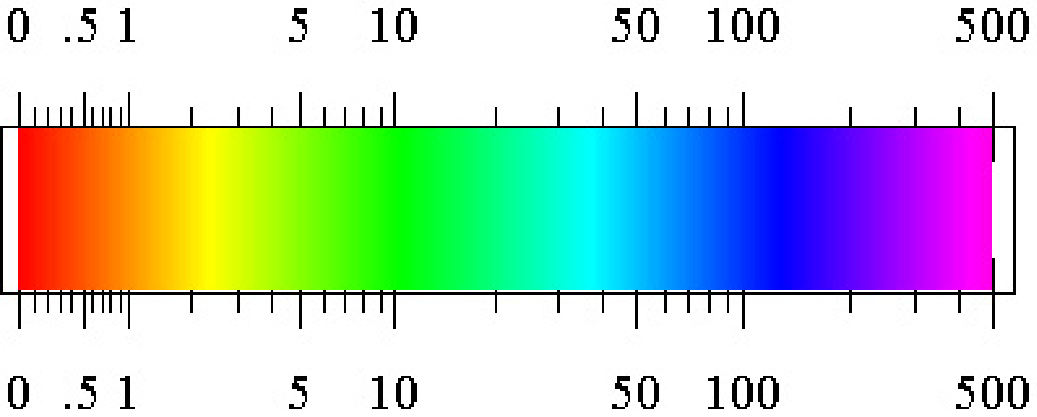}

	\begin{picture}(250,2)(0,0)
  \put(-21,4){(d)}
	\put(234,4){(e)}
	\end{picture}	
	
	\caption{(a), (b), (c) density and contour plots for lifetimes $\tau_1$, $\tau_2$, and $\tau_3$ of the lightest $N_1$, medium heavy $N_2$ and heaviest $N_3$ heavy neutrino respectively ($10^{-26}$~sec). The distances between two consecutive equipotential lines are $\Delta\tau_1=1\times10^{-26}$~sec for $\tau_1<10\times10^{-26}$~sec and for $\tau_1>10\times10^{-26}$~sec equipotential lines corresponding to $\tau_1=20\times10^{-26}$, $40\times10^{-26}$ and $200\times10^{-26}$~sec  are presented, $\Delta\tau_2=0.2\times10^{-26}$~sec, $\Delta\tau_3=0.1\times10^{-26}$~sec; (d)
	the detailed picture of the central part of  $\tau_1$;
$\Delta\tau_1=10\times10^{-26}$~sec for $\tau_1<100\times10^{-26}$~sec and $\Delta\tau_1=100\times10^{-26}$~sec for $\tau_1>100\times10^{-26}$~sec;
(e)
lifetime color coding.}  
	
\label{Lifetime}
\end{figure}

The resulting contour and density plots for masses, decay widths, and lifetimes of the heavy neutrinos $N_1$, $N_2$ and $N_3$ are presented in Figs.~\ref{Mass}--\ref{Lifetime}.
First, note that the graphs are symmetric under rotations by 
multiples of $2\pi/3$ as anticipated in section~\ref{symmetry}. 
Next, from Fig.~\ref{Mass}, we can easily see that the values of the heavy neutrino masses are larger than the $W$, $Z$, or Higgs thresholds for most of the parameter space,
justifying our use of two-body decay amplitudes. 
The mass of $N_1$ becomes smaller than these thresholds only in the vicinity of three isolated points at $r=1$, $\theta=-\pi/2+2\pi k/3$ ($k=0,1,2$), as illustrated in 
Fig.~\ref{Mass}d. 
As was shown in Ref.~\cite{okamura}, at these three points one of the $N$-fields 
is completely massless while the other two have degenerate mass. 
The lightest $N$ particle is completely stable at these points with zero decay width and infinite lifetime. 
However, the existence of such a light (less than $W$ and $Z$ thresholds) $N$ particle
is already ruled out experimentally by L3 \cite{Achard:2001qv} 
so we need not consider these points further.

It was also shown in Ref.~\cite{okamura} that at the center of the circle 
one heavy neutrino completely decouples from the light neutrino states 
(and therefore from the rest of the Standard Model particles) 
while the other two heavy states have degenerate masses. 
This decoupling can be seen in Fig.~\ref{Width}d and Fig.~\ref{Lifetime}d 
where at the center of the circle the decay width of the lightest heavy neutrino is zero 
and the lifetime is infinite. A similar decoupling occurs at the points where $r=1$ and $\theta=-5\pi/6+2\pi k/3$, $k=0,1,2$.

\begin{figure} 
	\centering
	\includegraphics[width=5cm]{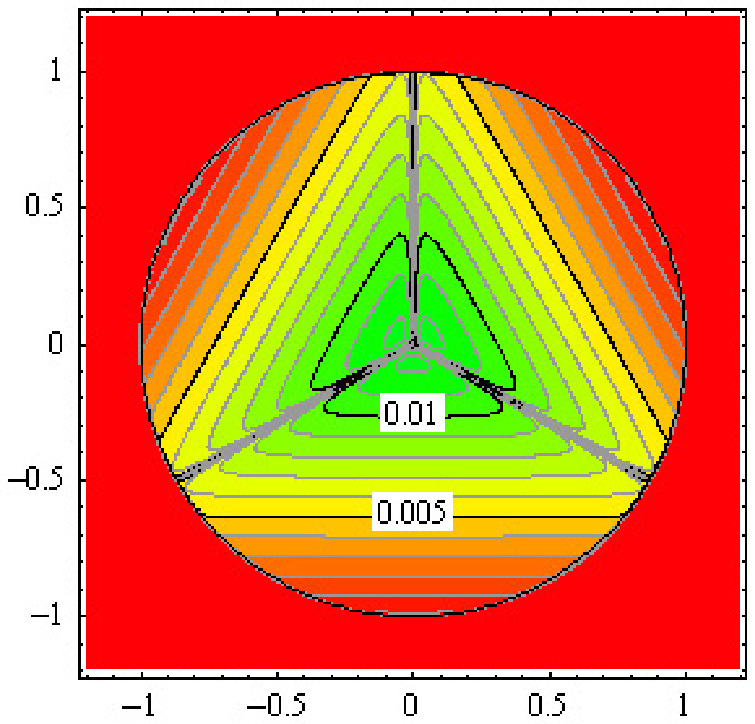}
	\includegraphics[width=5cm]{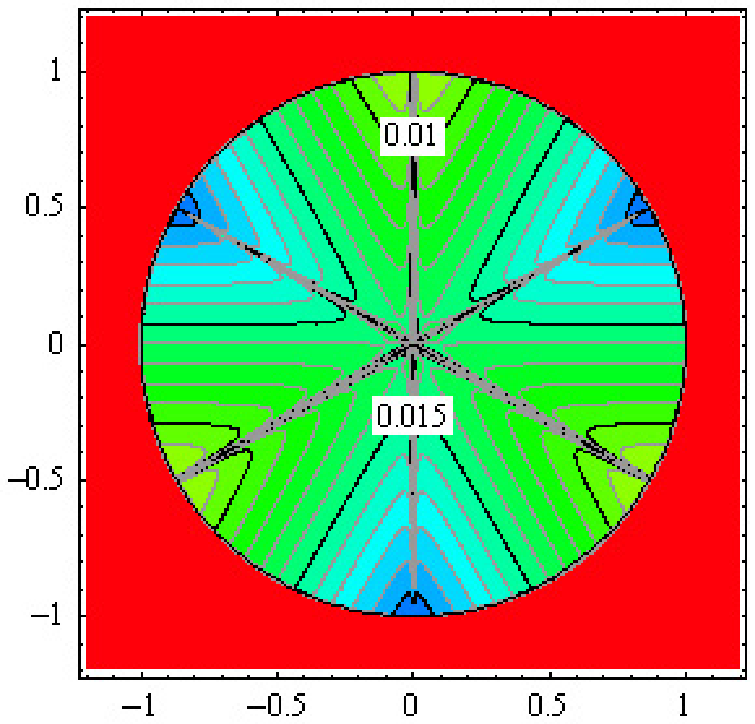}
	\includegraphics[width=5cm]{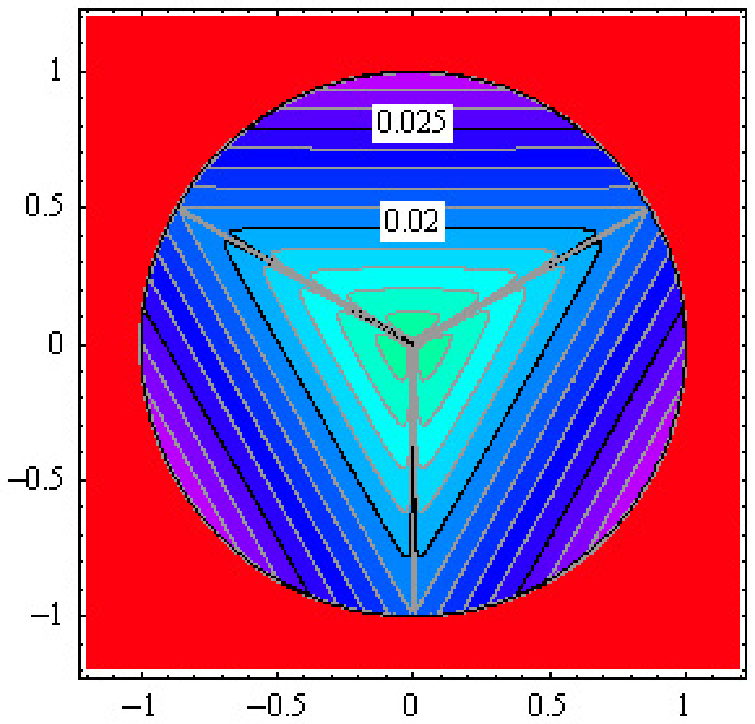}
	
	\begin{picture}(250,2)(0,0)
	\put(-21,4){(a)}
	\put(125,4){(b)}
	\put(271,4){(c)}
	\end{picture}

	\includegraphics[width=7cm]{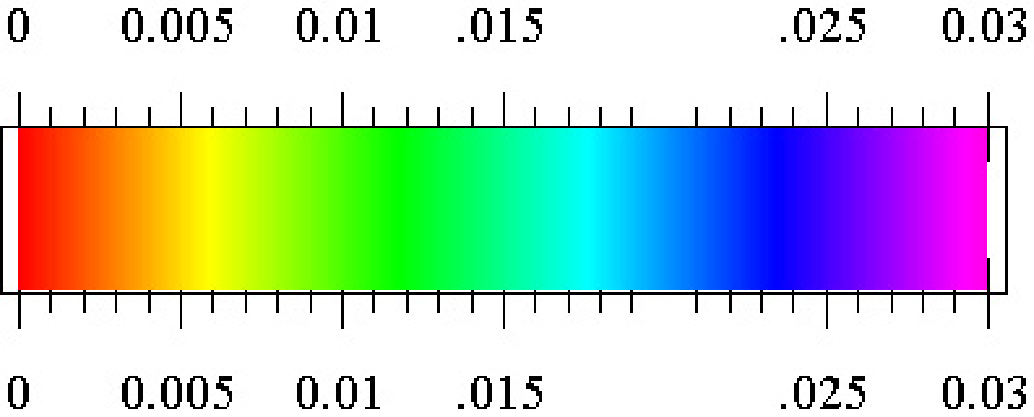}
	
	\begin{picture}(250,2)(0,0)
	\put(125,4){(d)}

	\end{picture}	
	
	\caption{(a), (b), (c) density and contour plots for the widths-to-mass ratios of the lightest $N_1$, medium heavy $N_2$ and heaviest $N_3$ heavy neutrinos respectively. The distances between two consecutive equipotential lines is $0.001$ for all plots; (d) mass-to-width ratio color coding.}  
	
\label{Ratios}
\end{figure}

Except for the vicinity of these points, the lifetimes of the $N$ particles 
are typically in the range of $10^{-26}$ to $10^{-24}$ seconds (see Fig.~\ref{Lifetime}).
Assuming that the particles are non-relativistic, the maximum distance they
can travel from their production points before decay is in the range of $10^{-17}$ to $10^{-15}$ meters.  If produced at colliders, they will decay inside the detector.
On the other hand,
the width-to-mass ratios of the particles are in the range of 0.1 to 3 percent as
shown in Fig.~\ref{Ratios}.
Therefore, the invariant mass spectrum of the decay products can be expected to show
a very narrow peak.

\section{Summary and Discussion}

In this paper, we calculate the mass spectrum, decay widths and lifetimes 
of the mostly-right-handed heavy neutrino states that appear in the model
proposed by Okamura et al. in Ref.~\cite{okamura}.
We map the parameter space of the Okamura model to the interior of a unit circle, 
and represent the results of our calculations as density-contour plots over it.
For the phenomenologically viable region of the model's parameter space, 
the heavy states have masses of a few TeV, and
are short-lived with the typical lifetimes from $10^{-26}$ to $10^{-24}$
seconds. At the same time, the decay widths are very small comparing to the masses. The typical width-to-mass ratio is in the range of $0.1$ to $3$ percent.

Though we have found that the heavy neutral particles in the Okamura model
have lifetimes in the range that allow for their observation at colliders, 
an analysis by Dicus, Karatas, and Roy \cite{Dicus:1991fk} suggests that
they may be difficult to observe at the LHC.
In Ref.~\cite{Dicus:1991fk}, the authors consider the production of like-sign leptons,
a lepton number violating process, 
as the signature of the heavy mostly-right-handed Majorana neutrino $N$:
this can occur through the process 
$W^\pm \rightarrow \ell^\pm N \rightarrow \ell^\pm\ell^\pm W^\mp 
\rightarrow \ell^\pm \ell^\pm+\mbox{jets}$, or
the $t$- and $u$-channel exchange of $N$ between 
two like-sign $W$'s radiated from the protons. 
Assuming masses in the range of $150\sim 4000$~GeV,
and a mixing as large as $\theta^2\sim 0.043$ (an order of magnitude larger than 
the Okamura model), the number of expected events at
the LHC was shown to be only a few per year; too few for them to be discernible above
the Standard Model background.

If the gauge group is extended to $SU(2)_L\times SU(2)_R\times U(1)_{B-L}$,
then the $N$'s can be copiously produced through the $W_R$ and $Z'$, as discussed 
in Refs.~\cite{Ho:1990dt,Datta:1992qw,ATLAS:1999fq}.
We will discuss the embedding of the Okamura texture into this gauge
structure in a subsequent paper \cite{PT2}.

\section*{Acknowledgments}

We would like to thank Naotoshi Okamura for helpful discussions, and
Kseniya Pronina for her help in the preparation of this paper. 
A portion of this work was first reported by Pronin at the 
Pheno 2003 Symposium, May 6, 2003, at the University of Wisconsin, Madison.
This research was supported by the U.S. Department of Energy, 
grant DE--FG05--92ER40709, Task A.


\end{document}